\documentclass{article}

\usepackage[utf8]{inputenc}
\usepackage[T1]{fontenc}
\usepackage[margin=1in]{geometry}
\usepackage{amsmath,amssymb,amsthm}
\usepackage{booktabs}
\usepackage{multirow}
\usepackage{algpseudocode}
\usepackage{hyperref}
\usepackage{xcolor}
\usepackage{authblk}
\usepackage{enumitem}

\hypersetup{
  colorlinks=true,
  linkcolor=blue!70!black,
  citecolor=blue!70!black,
  urlcolor=blue!70!black,
}

\newtheorem{remark}{Remark}

\newcommand{\hint}{\mathbf{h}}
\newcommand{\sz}{\mathbf{s}_2}
\newcommand{\sone}{\mathbf{s}_1}
\newcommand{\tzero}{\mathbf{t}_0}
\newcommand{\tone}{\mathbf{t}_1}
\newcommand{\boldA}{\mathbf{A}}
\newcommand{\boldw}{\mathbf{w}}
\newcommand{\boldz}{\mathbf{z}}
\newcommand{\boldy}{\mathbf{y}}
\newcommand{\wt}[1]{\mathrm{wt}(#1)}
\newcommand{\MakeHint}{\mathsf{MakeHint}}
\newcommand{\HighBits}{\mathsf{HighBits}}
\newcommand{\LowBits}{\mathsf{LowBits}}

\title{\textbf{The Hint Weight of ML-DSA Signatures Is Key-Dependent}\\[4pt]
\large An Empirical Study across the Three FIPS~204 Parameter Sets}

\author[1]{Dominik Blain}
\affil[1]{Ironproof \quad \texttt{dominik@ironproof.ai}}

\date{Ironproof-PQC Research Report IPQ-2026-01 \quad --- \quad May 2026, revised October 2026}

\begin{document}
\maketitle

\begin{abstract}
Every ML-DSA (FIPS~204) signature carries a public hint vector $\hint$.
We find that the Hamming weight of each hint polynomial $\hint_k$ depends on
the signing key: to first order it measures the Euclidean norm of
$(\tzero)_k$, the low-order part of $\mathbf{t}$ that key generation leaves out
of the public key.  A closed-form model predicts the per-key mean weight with
Pearson $r$ between $0.95$ and $0.98$; once the norm is accounted for, we
detect no key-dependent signal above sampling noise.

We measure the effect on the reference C implementation, with 200 keys and
2,000 signatures per key for each parameter set.  A one-way ANOVA rejects
key-independence of the total hint weight for ML-DSA-44, ML-DSA-65 and
ML-DSA-87 ($F = 30.1$, $10.3$, $11.1$).  The effect is weak: the key explains
0.5\% to 1.5\% of the variance of the total weight.  A single signature
identifies its key among 200 with $1.1$ to $1.3$ times chance accuracy from
the total weight, and $1.5$ to $1.8$ times from the per-polynomial weight
vector.  A one-sided test at significance $0.001$ separates two typical keys with
probability one half after about 1,300 to 3,700 signatures per key.

The hint weight does not endanger the signing key.  The Dilithium designers do not treat
$\tzero$ as secret, and $\tzero$ is known to be recoverable from signatures
\cite{azevedo2024uncompressing}.  The hint weight is, however, a weak
statistical fingerprint: with enough signatures, it can be used to test whether
they come from a common key, without the public key.  Bounded-weight signing, a backward-compatible filter
whose effect on the security argument we did not analyze, removes 86--91\% of
the between-key spread of the total weight but leaves the per-polynomial
channel largely intact.
\end{abstract}

\section{Introduction}
\label{sec:intro}

ML-DSA (formerly CRYSTALS-Dilithium~\cite{ducas2018crystals}) was standardized
by NIST in 2024 as FIPS~204~\cite{nist2024fips204}.  Each signature
$\sigma = (\tilde{c}, \boldz, \hint)$ contains a hint vector
$\hint \in \{0,1\}^{K \times N}$ that lets the verifier correct the rounding
error caused by compressing the public key.  The standard bounds the total
weight, $\wt{\hint} \le \omega$, but does not fix it.

\paragraph{Observation.}
The expected weight of each hint polynomial depends on the key.  Section~\ref{sec:model}
gives a first-order model in which
$\mathbb{E}[\wt{\hint_k}]$ is proportional to $\|(\tzero)_k\|_2$, and
Section~\ref{sec:l1} measures the effect on all three parameter sets.

\paragraph{Status of $\tzero$.}
$\tzero$ is stored in the ML-DSA private key, but it is not a secret on which
security rests.  The Dilithium specification states: ``The security of our
scheme does not rely on the part of the public key $\tzero$ being secret and so
we will be assuming that the public key is $\mathbf{t}$ rather than $\tone$''
\cite[Appendix~B]{dilithium2021spec}.  Azevedo Oliveira et al.\ show that each
signature leaks information on $\tzero$ and recover it in full from 200,000 to
500,000 signatures \cite{azevedo2024uncompressing}.  Berman et al.\ note
likewise that the hint test ``may leak information about $c\,\tzero$, but this
value is not secret'' \cite[Section~2.4]{berman2024note}.
Our observation is consistent with this line of work and does not contradict
the security claims of FIPS~204.

\paragraph{Practical relevance.}
A signature is normally verified under a known public key, so the signer is
not hidden.  The hint weight matters only where signatures are collected
without reliable key attribution: it lets an observer test whether two batches
of signatures come from the same key, using a few thousand signatures and only
the $K$ boundary bytes of each, without the public key.
ML-DSA makes no unlinkability claim, and we know of no deployed protocol that
relies on one.

\paragraph{Contributions.}
\begin{itemize}[leftmargin=*]
  \item A first-order model of the hint weight as a function of
        $\|(\tzero)_k\|_2$, checked against the reference implementation,
        with a residual analysis (Section~\ref{sec:model}).
  \item A measurement of the between-key effect on ML-DSA-44, ML-DSA-65 and
        ML-DSA-87, with effect sizes and sample complexity
        (Section~\ref{sec:l1}).
  \item A comparison of the per-polynomial weight vector with the total
        weight as an observable (Section~\ref{sec:l1vector}).
  \item Two negative results on countermeasures: fixed-iteration signing
        cannot change the hint-weight distribution, by argument
        (Section~\ref{sec:fis}), and
        bounded-weight signing suppresses the total-weight channel but not the
        per-polynomial one (Section~\ref{sec:bw}).
  \item A statement, under an explicitly idealized oracle, of what exact
        leakage of $c\,\tzero$ and $c\,\sone$ gives (Section~\ref{sec:oracle}).
\end{itemize}

\paragraph{What this paper does not claim.}
No key recovery from public signatures.  No weakness in Module-LWE or
Module-SIS.  No physical side-channel measurement: the oracle of
Section~\ref{sec:oracle} is simulated in software.  The limits of our data are
listed in Section~\ref{sec:limits}.

\section{Background}
\label{sec:background}

ML-DSA operates over $R_q = \mathbb{Z}_q[X]/(X^N+1)$ with $N = 256$,
$q = 8{,}380{,}417$, $d = 13$, and the parameters
$(K, L, \eta, \tau, \gamma_1, \gamma_2, \omega)$ of each security level.

\paragraph{Key generation.}
Sample $\sone \in R_q^L$ and $\sz \in R_q^K$ with coefficients in
$[-\eta, \eta]$, expand $\boldA \in R_q^{K \times L}$ from a public seed, and
compute $\mathbf{t} = \boldA \sone + \sz$.  Split
$\mathbf{t} = 2^d\,\tone + \tzero$ with
$\tzero \in (-2^{d-1}, 2^{d-1}]^{KN}$.  The public key holds $\tone$; the private
key holds $(\sone, \sz, \tzero)$, together with $\rho$, a private seed and a hash of the
public key.

\paragraph{Signing.}
Repeat until all checks pass:
\begin{align*}
  \boldy &\leftarrow R_q^L,\quad \|\boldy\|_\infty \le \gamma_1, \qquad
  \boldw = \boldA \boldy,\\
  c &= \mathsf{SampleInBall}\big(H(\mu \,\|\, \HighBits(\boldw))\big), \qquad
  \boldz = \boldy + c\,\sone,\\
  \hint &= \MakeHint(-c\,\tzero,\; \boldw - c\,\sz + c\,\tzero).
\end{align*}
The challenge $c$ has exactly $\tau$ coefficients in $\{-1,+1\}$ and is zero
elsewhere.  A signing attempt is rejected if $\|\boldz\|_\infty$ or the low
part of $\boldw - c\,\sz$ is too large, if $\|c\,\tzero\|_\infty \ge \gamma_2$,
or if $\wt{\hint} > \omega$.

\paragraph{The hint.}
$\hint_{k,j} = 1$ exactly when adding $(c\,\tzero)_{k,j}$ to
$(\boldw - c\,\sz)_{k,j}$ changes its high bits.  The bound is
$\omega = 80$ (ML-DSA-44), $55$ (ML-DSA-65), $75$ (ML-DSA-87).

\section{A First-Order Model of the Hint Weight}
\label{sec:model}

Write $r = (\boldw - c\,\sz)_{k,j}$ and $u = (c\,\tzero)_{k,j}$.  The low part
$r_0$ of $r$ lies in $(-\gamma_2, \gamma_2]$ and is close to uniform.  For
$|u| < \gamma_2$, adding $u$ changes the high bits of $r$ when $r_0 + u$ leaves
that interval, which happens with probability about $|u| / (2\gamma_2)$.
The coefficient $u$ is a signed sum of $\tau$ coefficients of $(\tzero)_k$; over
random challenges it is approximately Gaussian with variance
$\tau\,\|(\tzero)_k\|_2^2 / N$, so
$\mathbb{E}|u| \approx \sqrt{2/\pi}\,\sqrt{\tau/N}\,\|(\tzero)_k\|_2$.
Summing over the $N$ coefficients:
\begin{equation}
\label{eq:model}
  \mathbb{E}\big[\wt{\hint_k}\big] \;\approx\;
  \frac{\sqrt{2\tau N/\pi}}{2\gamma_2}\;\|(\tzero)_k\|_2 .
\end{equation}

The model ignores the conditioning on acceptance and the truncation at
$\omega$; it is a heuristic, not a theorem.  For a key with uniform $\tzero$
it predicts a total weight of $63.4$, $38.7$ and $57.1$ for the three
parameter sets.

\paragraph{Check.}
For each of the 200 keys of Section~\ref{sec:l1} we compute
$\|(\tzero)_k\|_2$ from the private key and compare the prediction of
\eqref{eq:model} with the mean weight observed over 2,000 signatures.

\begin{center}
\begin{tabular}{lcccc}
\toprule
\textbf{Variant} & \textbf{$r$ per polynomial} & \textbf{$r$ per key (total)} &
\textbf{Predicted total} & \textbf{Observed total} \\
\midrule
ML-DSA-44 & $0.981$ & $0.983$ & $63.33$ & $62.76$ \\
ML-DSA-65 & $0.958$ & $0.955$ & $38.67$ & $38.12$ \\
ML-DSA-87 & $0.956$ & $0.960$ & $57.21$ & $56.66$ \\
\bottomrule
\end{tabular}
\end{center}

\noindent
The model overestimates the mean by 0.9\% to 1.4\%.

\paragraph{Residuals.}
A correlation below one does not by itself show a second effect: each observed
mean carries sampling noise of variance $\sigma_k^2/2000$.  We regress the
observed means on the prediction and compare the residuals with that noise.

\begin{center}
\begin{tabular}{lcccc}
\toprule
 & \multicolumn{3}{c}{\textbf{Variance of the per-polynomial means}} & \\
\cmidrule(lr){2-4}
\textbf{Variant} & \textbf{Total} & \textbf{Residual} & \textbf{Sampling noise} &
\textbf{$\chi^2$ test of the residuals} \\
\midrule
ML-DSA-44 & $0.2021$ & $0.0077$ & $0.0072$ & $856.4$ on $798$ d.f., $p = 0.07$ \\
ML-DSA-65 & $0.0339$ & $0.0028$ & $0.0031$ & $1084.9$ on $1198$ d.f., $p = 0.99$ \\
ML-DSA-87 & $0.0413$ & $0.0036$ & $0.0034$ & $1663.2$ on $1598$ d.f., $p = 0.13$ \\
\bottomrule
\end{tabular}
\end{center}

\noindent
The residual variance is at the level of the sampling noise for all three
parameter sets, and the $\chi^2$ test does not reject the hypothesis that the
residuals are noise only.  With 2,000 signatures per key we detect no
key-dependent signal beyond the norm.  The key-dependent quantity observed
through the hint weight is consistent, to the precision of this experiment, with
the vector of norms alone, $(\|(\tzero)_0\|_2, \ldots, \|(\tzero)_{K-1}\|_2)$.

\section{Measurements}
\label{sec:l1}

\subsection{Data}
\label{sec:data}

The adversary sees valid signatures only.  The per-polynomial weights are read
from the $K$ boundary bytes at the end of the hint encoding; the last of them
is the total weight.  No timing, no interaction with the signer.

All measurements use the pq-crystals reference C implementation (\texttt{ref/},
commit \texttt{6e00625}), unmodified except for the random source, which we
replace by a SHAKE256 stream so that a run can be repeated bit for bit.  For
each parameter set we generate 200 keys and 2,000 signatures per key on
distinct messages: 400,000 signatures per parameter set, 1,200,000 in all.
The first 20 signatures of every key are checked with the reference verifier.
The same keys then produce 2,000 signatures each under bounded-weight signing
(Section~\ref{sec:bw}).  The code was built with Apple clang~17.0.0 at
\texttt{-O2} and run on an Apple M4 under macOS~26.2.

An earlier, independent run of the same code for ML-DSA-44 (50 keys, 2,000
signatures per key, GCC~13.3.0, Linux~6.8.0 on aarch64, system randomness)
gives the same picture: mean $62.88$, within-key standard deviation $7.32$,
$F = 25.07$, $\eta^2 = 0.012$.

\subsection{Total Weight}

\begin{table}[h!]
\centering
\caption{Total hint weight, 200 keys $\times$ 2,000 signatures.  $\sigma_w$ is
the pooled within-key standard deviation, ``SD of means'' the standard
deviation of the 200 per-key means, ``spread'' the difference between the
largest and smallest per-key mean.  ANOVA degrees of freedom are
$(199,\ 399{,}800)$.}
\label{tab:l1}
\begin{tabular}{lrrrrrrr}
\toprule
\textbf{Variant} & \textbf{Mean} & $\sigma_w$ & \textbf{SD of means} &
\textbf{Spread} & $F$ & $p$ & $\eta^2$ \\
\midrule
ML-DSA-44 & 62.758 & 7.344 & 0.902 & 4.397 & 30.14 & $< 10^{-300}$ & 0.0148 \\
ML-DSA-65 & 38.124 & 5.995 & 0.430 & 2.974 & 10.30 & $< 10^{-300}$ & 0.0051 \\
ML-DSA-87 & 56.661 & 7.201 & 0.535 & 2.554 & 11.05 & $< 10^{-300}$ & 0.0055 \\
\bottomrule
\end{tabular}
\end{table}

The key explains between 0.5\% and 1.5\% of the variance of the total weight
($\eta^2$).  The very small $p$-values reflect the sample size of 400,000
signatures per variant, not the strength of the effect.  A permutation test on
the key labels (1,000 permutations per variant) never produced an $F$ as large
as the observed one ($p < 1/1001$).

\paragraph{Sample complexity.}
Table~\ref{tab:samples} gives three figures, all at significance $0.001$:
the number of signatures per key after which the 200-key ANOVA rejects
key-independence; and, for the two most extreme keys of the dataset and for two
keys whose means differ by one standard deviation of the per-key means, the
number per key at which a one-sided two-sample $z$-test separates them with
probability one half, $n = 2\,(z_{0.999}\,\sigma_w/\Delta)^2$.  For 80\% power
the last two figures grow by a factor of about $1.6$.  The
extreme pair is selected after the fact and is a best case, not a typical one.

\begin{table}[h!]
\centering
\caption{Signatures per key at significance $0.001$ (pairs: one-sided test, power $0.5$).}
\label{tab:samples}
\begin{tabular}{lrrr}
\toprule
\textbf{Variant} & \textbf{200-key ANOVA} & \textbf{Extreme pair} &
\textbf{Pair one SD apart} \\
\midrule
ML-DSA-44 & 26 & 53  & 1{,}267 \\
ML-DSA-65 & 66 & 78  & 3{,}707 \\
ML-DSA-87 & 36 & 152 & 3{,}456 \\
\bottomrule
\end{tabular}
\end{table}

\subsection{Per-Polynomial Weight Vector}
\label{sec:l1vector}

The total weight collapses $K$ polynomials into one integer.
Equation~\eqref{eq:model} predicts that each $\wt{\hint_k}$ carries its own
key-dependent signal, since the $K$ norms of a key are, to a good approximation, independent.

\paragraph{Per-coordinate ANOVA.}
Every coordinate is key-dependent on its own.  The per-coordinate $F$ lies
between $25.1$ and $31.4$ for ML-DSA-44, between $9.6$ and $13.1$ for
ML-DSA-65, and between $11.0$ and $13.3$ for ML-DSA-87; every $p$-value is
below $10^{-270}$.  Cohen's $d$ for the most extreme key pair of a coordinate
ranges from $0.67$ to $0.72$, from $0.37$ to $0.51$, and from $0.37$ to
$0.47$.

\paragraph{Covariance.}
The pooled within-key covariance for ML-DSA-44 is nearly diagonal:
\[
\hat{\Sigma}_\text{within} \approx
\begin{pmatrix}
 14.46 & -0.32 & -0.29 & -0.31 \\
-0.32 &  14.44 & -0.34 & -0.31 \\
-0.29 & -0.34 &  14.44 & -0.35 \\
-0.31 & -0.31 & -0.35 &  14.43
\end{pmatrix}.
\]
The small negative off-diagonal terms are consistent with the bound
$\wt{\hint} \le \omega$: a high count in one polynomial slightly
suppresses the others.  For ML-DSA-65 and ML-DSA-87 the off-diagonal terms
are at most $0.07$ in absolute value against a diagonal of $6.2$ and $6.8$.

\paragraph{Classification.}
We assign a single signature to one of the 200 keys, training on the first
1,000 signatures of each key and testing on the last 1,000.  The baseline is
$0.5\%$.  A nearest-mean classifier on the total weight is compared with a
nearest-centroid classifier on the weight vector, in Mahalanobis distance with
the pooled within-key covariance (Table~\ref{tab:multivarlift}).  A two-sample
Hotelling $T^2$ test on 100 randomly drawn key pairs, with 2,000 signatures
per key, separates 99\%, 97\% and 100\% of the pairs at $\alpha = 0.05$.

\begin{table}[h!]
\centering
\caption{Single-signature 200-way classification: lift over the 0.5\%
baseline.}
\label{tab:multivarlift}
\begin{tabular}{lcccc}
\toprule
\textbf{Variant} & $K$ & \textbf{Total weight} & \textbf{Weight vector} &
\textbf{$T^2$-separated pairs} \\
\midrule
ML-DSA-44 & 4 & $1.27$ & $1.76$ & 99\% \\
ML-DSA-65 & 6 & $1.23$ & $1.49$ & 97\% \\
ML-DSA-87 & 8 & $1.12$ & $1.62$ & 100\% \\
\bottomrule
\end{tabular}
\end{table}

The vector gives a higher lift than the total weight, as the model predicts: two keys can have
the same total weight and different per-polynomial norms.  The channel remains
narrow.  We did not estimate the mutual information between the weight vector
and the key, so we cannot say how far a better classifier could go.

\section{Fixed-Iteration Signing and the Hint Weight}
\label{sec:fis}

Fixed-iteration signing runs a constant number of signing attempts and selects
the first valid one in constant time.  It removes the dependence of signing
time on the number of rejections.  It cannot change the distribution of the
hint: the selected attempt is an ordinary accepted attempt, and the selection
criterion is validity, not hint weight.  This is an argument, not a
measurement: we did not simulate a fixed-iteration signer.

\section{What an Idealized Oracle on the Ring Products Would Give}
\label{sec:oracle}

This section is not an attack on any implementation.  It records what follows
if an adversary learns exact coefficients of the products computed during
signing, which is the target of the physical attacks cited below.

\paragraph{Oracle.}
For a signing attempt with challenge $c$, the oracle returns every coefficient
of $c\,(\tzero)_k$ and of $c\,(\sone)_l$, without noise.  Profiled power and
electromagnetic attacks recover such intermediate values partially and with
noise \cite{ravi2018side,berzati2023exploiting,marzougui2022profiling,qiao2023ntt}.
Turning noisy partial leakage into the exact oracle assumed here is the hard
part of those works and is not addressed by ours.

\paragraph{Linear algebra.}
Multiplication by $c$ in $R_q$ is a linear map with matrix $M(c)$, the
negacyclic convolution matrix of $c$.  When $M(c)$ is invertible over
$\mathbb{Z}_q$, one oracle answer determines the operand:
$(\tzero)_k = M(c)^{-1}\,\bigl(c\,(\tzero)_k\bigr)$, and the same for
$(\sone)_l$.  Then $\sz = 2^d\,\tone + \tzero - \boldA\,\sone$.

\paragraph{Simulation.}
With keys and oracle answers produced by a Python model
(\texttt{dilithium\_sim.py}) at ML-DSA-44 parameters, one oracle answer on $c\,\tzero$ yields all 1,024 coefficients of
$\tzero$ in under 20~s of exact Gaussian elimination in Python (script
\texttt{recover\_t0\_real.py}).  Adding one answer on $c\,\sone$ yields all
3,072 coefficients of $(\tzero, \sone, \sz)$ in under 40~s (script
\texttt{full\_key\_recovery\_real.py}).  The values are accepted without
reference to the key and then compared with it: they equal the generated key.
The key also signs: with a fresh random seed in place of the secret seed, 20 of
20 signatures made with it are accepted by a verifier that uses only the public
key, which accepts 20 of 20 honest signatures and rejects 20 of 20 signatures
made with a key differing in one coefficient (script
\texttt{verify\_recovered\_key.py}).

This does not touch Module-LWE: the oracle hands over the secret through a
linear map.  It shows only that exact access to $c\,\tzero$ and $c\,\sone$ at one
signing attempt determines the key.

\begin{remark}[The rejection test on $c\,\tzero$]
The test $\|c\,\tzero\|_\infty \ge \gamma_2$ can fire only for ML-DSA-44:
$\tau\,2^{d-1}$ equals $159{,}744 > \gamma_2 = 95{,}232$ there, and
$200{,}704$ and $245{,}760$, both below $\gamma_2 = 261{,}888$, for ML-DSA-65
and ML-DSA-87.  Even for ML-DSA-44 it almost never fires: the coefficients of
$c\,\tzero$ have standard deviation about $14{,}800$, more than six standard
deviations below $\gamma_2$.  The Dilithium specification considers it safe to
reveal which rejection test failed and which coefficient violated the bound
\cite[Section~5.5]{dilithium2021spec}; we read this as consistent with $\tzero$
not being treated as secret.
\end{remark}

\subsection{A Passive Estimate of $\tzero$}
\label{sec:passive}

From public data the verifier computes
$\boldw' = \boldA\boldz - c\,\tone\,2^d = \boldw - c\,\sz + c\,\tzero$.  At
positions without a hint bit,
$\LowBits(\boldw')_{k,j} = (w_0)_{k,j} - (c\,\sz)_{k,j} + (c\,\tzero)_{k,j}$,
where $w_0$ is the low part of $\boldw$.  Correlating these low bits with the
challenge over $M$ signatures,
\[
  R_k[m] = \sum_{i=1}^{M}\sum_{j=0}^{N-1}
  (1-h_{i,k,j})\;\LowBits(\boldw'_i)_{k,j}\; c_i[(j-m) \bmod N]
\]
(with the sign of the negacyclic wrap-around), gives an estimate of
$(\tzero)_k[m]$ up to a scale factor.

We ran this with \texttt{dilithium\_sim.py} at ML-DSA-44 parameters and
$M = 3{,}000$ signatures (script \texttt{l2d\_differential\_test.py}).  The
estimate has Pearson correlation of $0.99$ with the true $(\tzero)_k$ for each of
the four polynomials.  With the $\tzero$ of an unrelated key the correlations are
below $0.07$ in absolute value, consistent with zero (for $N = 256$ the null
standard deviation is about $1/\sqrt{256} \approx 0.06$).  Signing is randomized,
so the values vary slightly between runs.  This simple estimator does not reach exact coefficients.
Azevedo Oliveira et al.\ \cite{azevedo2024uncompressing} do, with a finer
method and 200,000 to 500,000 signatures; our estimate, obtained in
simulation, is an independent and cruder confirmation of the leakage they
establish.

\section{Mitigations}
\label{sec:mitigations}

\subsection{Bounded-Weight Signing}
\label{sec:bw}

Injecting dummy hint bits is not possible: a hint bit at a position without a
carry makes verification fail.  A signer can, however, discard signatures
whose total hint weight falls outside a fixed public window and sign again.

\begin{algorithmic}[1]
\Function{BW-Sign}{$\mathit{sk}, \mu, \ell, u$}
  \Repeat
    \State $(\tilde{c}, \boldz, \hint) \gets \text{ML-DSA.Sign}(\mathit{sk}, \mu)$
           \Comment{fresh randomness}
  \Until{$\ell \le \wt{\hint} \le u$}
  \State \Return $(\tilde{c}, \boldz, \hint)$
\EndFunction
\end{algorithmic}

The output verifies under unmodified FIPS~204.  Our harness signs a fresh
message at each attempt rather than the same $\mu$; with randomized signing this
does not change the distribution of the hint weight.  We use a window of half-width
3 around the rounded grand mean: $[60, 66]$, $[35, 41]$ and $[54, 60]$, and
measure it on the reference implementation with the same 200 keys.

\begin{table}[h!]
\centering
\caption{With and without bounded-weight signing, 200 keys $\times$ 2,000
signatures.  ``Spread'' and $F$ refer to the total weight; the lifts are
single-signature 200-way classification lifts; ``cost'' is the mean number of
full signing runs per output signature.}
\label{tab:bwresults}
\begin{tabular}{llrrrrr}
\toprule
\textbf{Variant} & \textbf{Setting} & \textbf{Spread} & $F$ &
\textbf{Total lift} & \textbf{Vector lift} & \textbf{Cost} \\
\midrule
\multirow{2}{*}{ML-DSA-44}
  & Standard          & 4.40 & 30.14 & $1.27$ & $1.76$ & $1$ \\
  & BW-Sign $[60,66]$ & 0.41 &  3.35 & $1.08$ & $1.69$ & $2.83$ \\
\midrule
\multirow{2}{*}{ML-DSA-65}
  & Standard          & 2.97 & 10.30 & $1.23$ & $1.49$ & $1$ \\
  & BW-Sign $[35,41]$ & 0.35 &  2.07 & $1.03$ & $1.46$ & $2.30$ \\
\midrule
\multirow{2}{*}{ML-DSA-87}
  & Standard          & 2.55 & 11.05 & $1.12$ & $1.62$ & $1$ \\
  & BW-Sign $[54,60]$ & 0.36 &  1.90 & $1.01$ & $1.63$ & $2.75$ \\
\bottomrule
\end{tabular}
\end{table}

Bounded-weight signing removes 86\% to 91\% of the between-key spread of the
total weight and brings the total-weight classifier close to chance.  It does
not remove the effect: the ANOVA on the total weight still rejects
key-independence ($p = 1\times10^{-51}$, $6\times10^{-17}$ and $3\times10^{-13}$).
The per-polynomial channel is largely unaffected: the per-coordinate $F$ values stay in
the same range as without the filter, and the vector lift does not move beyond
what the test split can resolve.  Keys with the same total weight still
distribute it differently across their $K$ polynomials.

The cost is 2.3 to 2.8 signing runs per signature.  We did not analyze the
effect of the extra rejection on the zero-knowledge argument of the scheme;
the filter should not be deployed on the strength of this paper alone.

\subsection{Constant Per-Polynomial Weight}
\label{sec:cwencode}

Removing both channels requires the weight of each hint polynomial to be
independent of the key.  We do not expect filtering to achieve this at
acceptable cost; it would likely need a change in how the hint is encoded and used, that is, a
change to the standard.  We have not designed or evaluated such an encoding.

\subsection{Key Rotation}

Where linkability matters, limiting the number of signatures issued under one
key to well below the figures of Table~\ref{tab:samples} limits what the hint
weight reveals.  It does nothing against an observer who holds the public key.

\section{Related Work}
\label{sec:related}

\paragraph{Leakage of $\tzero$.}
The closest work is Azevedo Oliveira, Calle Viera, Cogliati and Goubin
\cite{azevedo2024uncompressing}, who show that each signature leaks
information on $\tzero$ and recover it from 200,000 to 500,000 signatures.
They note that this does not compromise the security proof but makes
side-channel attacks more efficient.  Wang, G\"artner and Dubrova \cite{wang2024decompressing} partially recover
the least significant bit of $\tzero$ from power traces with profiled deep
learning, which roughly halves the number of signatures that reconstruction
needs.  Berman, Doubchak and Livne
\cite{berman2024note} refine the test that precedes the hint computation and
observe that it may leak information about $c\,\tzero$, which is not secret.
We add a different view of the same leakage: a one-byte statistic per
polynomial, its first-order explanation, and its measured strength.

\paragraph{Physical attacks on the products.}
Ravi et al.\ \cite{ravi2018side}, Berzati et al.\ \cite{berzati2023exploiting},
Marzougui et al.\ \cite{marzougui2022profiling} and Qiao et al.\
\cite{qiao2023ntt} recover Dilithium key material from power or
electromagnetic leakage of intermediate values, with real measurements and
noise.  Azouaoui et al.\ \cite{azouaoui2023protecting} study which
intermediates need protection.  Section~\ref{sec:oracle} assumes the outcome
of such attacks in an idealized form and adds nothing to their feasibility.

\paragraph{Other attacks on the signing procedure.}
Krahmer et al.\ \cite{krahmer2024correction} use fault injection against
randomized Dilithium.  Azevedo Oliveira, Beraud and Goubin
\cite{azevedo2025implicit} recover the key when the nonces of several
signatures share information.  Both are unrelated to the hint weight.

\section{Limitations}
\label{sec:limits}

\begin{itemize}[leftmargin=*]
  \item \textbf{One implementation.}  The hint-weight measurements use the
        reference C implementation only.  The hint is part of the signature
        format, so a conformant implementation should behave the same, but we
        did not measure another one.
  \item \textbf{Number of keys.}  200 keys per parameter set.  The extreme
        pairs and the spread depend on this sample.
  \item \textbf{Model.}  Equation~\eqref{eq:model} is heuristic.  The residual
        analysis has the resolution of 2,000 signatures per key; a smaller
        second-order effect would go unseen.
  \item \textbf{Simulations.}  Section~\ref{sec:oracle} relies on a Python
        model of the signing algorithm that is not byte-compatible with
        FIPS~204 (hash inputs and encodings differ; $\mathsf{Decompose}$ and the
        challenge sampler follow the standard).  The oracle is exact and
        noise-free; no device was measured.
        Section~\ref{sec:fis} is an argument, without simulation.
  \item \textbf{Mitigation.}  Bounded-weight signing was evaluated for its
        effect on the hint weight only, not for its effect on the security
        argument of the scheme.
  \item \textbf{Classifier.}  Nearest-centroid classifiers are simple; no
        bound on the information carried by the weight vector is given.
\end{itemize}

\section{Conclusion}
\label{sec:conclusion}

The weight of each hint polynomial in an ML-DSA signature is a noisy
measurement of the norm of the corresponding polynomial of $\tzero$; with
2,000 signatures per key we find nothing else in it.  In the reference
implementation the effect is present on all three parameter sets and is weak: 0.5\% to 1.5\% of the variance of the
total weight, a single-signature classification lift of at most $1.8$ among
200 keys, and thousands of signatures to separate two typical keys.  It does
not threaten the signing key and is consistent with the known fact that
$\tzero$ leaks from signatures.  The practical consequence we identify is a
weak linkability fingerprint.  Fixed-iteration signing cannot change it
(Section~\ref{sec:fis}); filtering on the total weight reduces the
total-weight channel and leaves the per-polynomial one; we did not evaluate a
filter on the per-polynomial weights.

\paragraph{Reproducibility.}
The measurement harness (\texttt{hint\_collect.c}, with its deterministic
random source and \texttt{run\_all.sh}), the raw per-polynomial weights of the
2,400,000 signatures, and the analysis scripts are available at
\url{https://github.com/dom-omg/ml-dsa-hint-weight}.
The script \texttt{paper\_numbers.py} recomputes every figure of
Sections~\ref{sec:model}, \ref{sec:l1} and \ref{sec:bw} from the raw files.

\section*{Acknowledgments}

This work uses the pq-crystals Dilithium reference implementation
(\url{https://github.com/pq-crystals/dilithium}), NumPy~\cite{harris2020array}
and SciPy.

\begingroup\sloppy\hbadness=10000

\endgroup

\end{document}